\documentclass[10pt,journal]{IEEEtranTCOM}

\usepackage[english]{babel}
\usepackage[usenames]{color}
\usepackage[cp1250]{inputenc}
\usepackage{amsfonts}
\usepackage{amsthm}
\usepackage{graphicx}
\usepackage{epsfig}
\usepackage{mathrsfs}
\usepackage{amsmath}
\usepackage{algorithm}
\usepackage{algorithmic}

\theoremstyle{plain}

\begin{document}
\newcommand{\bea}{\begin{eqnarray}}
\newcommand{\eea}{\end{eqnarray}}
\newcommand{\be}{\begin{equation}}
\newcommand{\ee}{\end{equation}}
\newcommand{\beas}{\begin{eqnarray*}}
\newcommand{\eeas}{\end{eqnarray*}}
\newcommand{\bs}{\backslash}
\newcommand{\bc}{\begin{center}}
\newcommand{\ec}{\end{center}}
\def\SC {\mathscr{C}}

\title{Modeling joint probability distribution \\ of yield curve parameters}
\author{\IEEEauthorblockN{Jarek Duda$^1$} \IEEEauthorblockN{Ma{\l}gorzata Snarska$^2$}\\
\IEEEauthorblockA{$^1$ Jagiellonian University,
Golebia 24, 31-007 Krakow, Poland,
Email: \emph{dudajar@gmail.com}}\\
\IEEEauthorblockA{$^1$ Cracow University of Economics, Faculty of Finance and Law,
Rakowicka 27, 31-510 Krakow, Poland,
Email: \emph{snarskam@uek.krakow.pl}}\\
}
\maketitle

\begin{abstract}
US Yield curve has recently collapsed to its most flattened level since subprime crisis and is close to the inversion. This fact has gathered attention of investors around the world and revived the discussion of proper modeling and forecasting yield curve, since changes in interest rate structure are believed to represent investors expectations about the future state of economy and in this sense have foreshadowed each and every single recession in  the United States. Although changes in term structure of interest rates are relatively easy to interpret they are however very difficult to model and forecast due to no proper economic theory underlying such events. Yield curves are usually represented by multivariate yet quite sparse time series ie.  at any point in time infinite dimensional curve is portrayed via relatively few points in a multivariate space of data and as a consequence multimodal statistical dependencies behind these curves are relatively hard to extract and forecast via typical multivariate statistical methods.
We propose to model yield curves via reconstruction of joint probability distribution of parameters in functional space as a high degree polynomial. Thanks to adoption of an orthonormal basis, the MSE estimation of coefficients of a given function is just an average over a data sample in the space of functions. 
Since such polynomial coefficients are independent and have cumulant-like interpretation: ie. they describe corresponding perturbation from an uniform joint distribution, our approach can also be extended to any $d$-dimensional space of yield curve parameters (also in neighboring times) due to controllable accuracy. We believe that this approach to modeling of local behavior of a sparse multivariate curved time series can complement prediction from standard models like ARIMA, that are using long range dependencies, but provide only inaccurate prediction of probability distribution, often as just Gaussian with constant width.
\end{abstract}
\textbf{Keywords:} time series analysis, machine learning, density estimation, yield curve modeling.
\section{Introduction}
As successful forecasting of financial time series could be often turned into profit, it makes it difficult to get a better prediction for the following value that just the previous value. However, above self-regulatory mechanism of the market does not restrict prediction of probability distribution of values, what is crucial for example for risk evaluation or Monte Carlo simulations.

\begin{figure}[t!]
    \centering
        \includegraphics{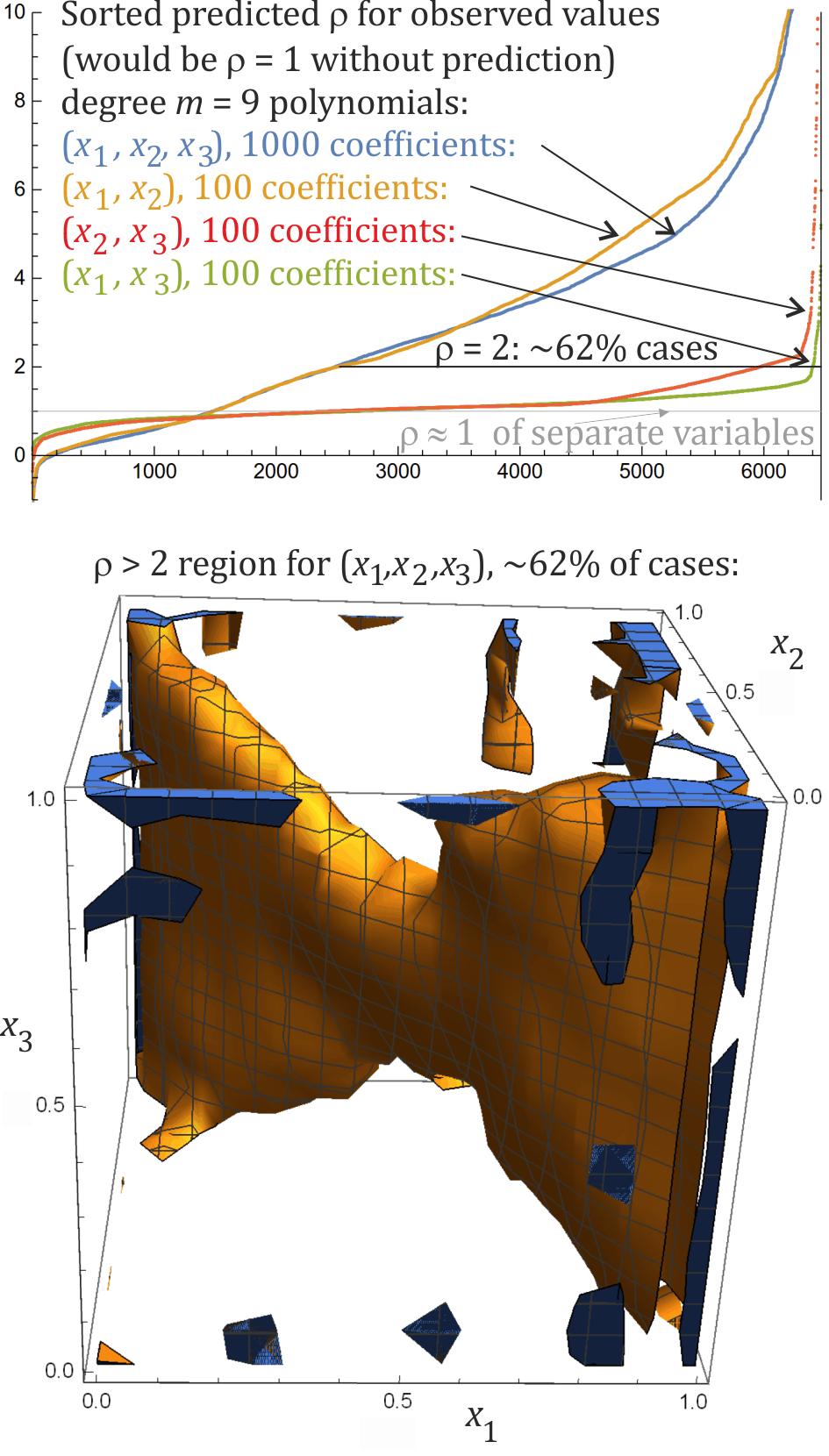}
        \caption{Each variable is normalized to have nearly uniform $\rho=1$ density (PDF) on $[0,1]$ range. Top: sorted predicted $\rho$ for observed values is usually much higher than base $\rho=1$ thanks to exploiting joint distribution. Four graphs correspond to joint distribution of parameters reconstructed with degree 9 polynomial for all three variables, or all their pairs. Surprisingly, we see that $(x_1, x_2)$ gives even better predictions than for all 3 variables here. Bottom: region of predicted $\rho>2$ for all 3 variables. From above plot we can read that observed values were there in $\approx 62 \%$ of cases. In contrast to usually assumed Gaussian, distribution obtained from the real data turns out multimodal here. Density focused near diagonal for $(x_1,x_2)$ means they are anti-correlated.}
       \label{summ}
\end{figure}
Standard approaches to predict probability distribution of values like ARIMA usually models this distribution as Gaussian, often of constant width: predicts some value and its inaccuracy (standard deviation). In contrast, we will model this probability distribution using large number of independent coefficients describing joint distribution as polynomial - what turns out leading to very different and more complex distribution than standardly assumed Gaussian - for example multimodal in the discussed example.

Specifically, as it is difficult to obtain a better prediction than just the previous value, we will focus on sequence of differences between two succeeding values. In discussed example it will be 3-dimensional space of parameters of Yield Curves of Diebold-Li model~\cite{diebold} (for fixed $\lambda=0.0609$), which dimensionality can be further increased for improved prediction by operating on time window: use a few previous values as context for prediction.

\begin{figure}[t!]
    \centering
        \includegraphics{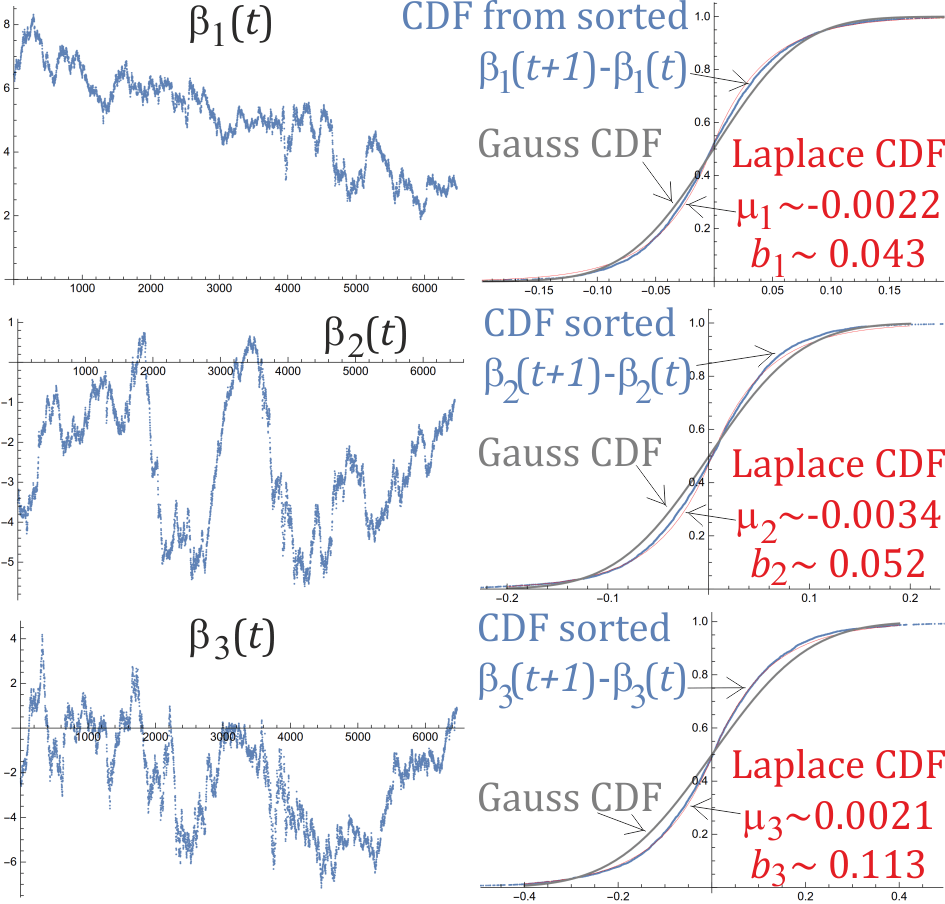}
        \caption{Left: time series of 6470 (from 1993 to 2018) daily Yield Curve $\beta_1, \beta_2, \beta_3$ parameters (Diebold-Li model~\cite{diebold}) fitted using $\lambda=0.0609$ standard assumption. We will work on $x_i(t):=CDF_{Laplace(\mu_i,b_i)}(\beta_i(t+1)-\beta_i(t))$ time series: normalized to nearly uniform distribution on $[0,1]$. Right: comparison of empirical CDF obtained from sorted values with CDF of Laplace and Gaussian distribution with estimated parameters - we will use Laplace as it has better agreement.}
       \label{norm}
\end{figure}

For convenience of fitting polynomial, we will first normalize each variable to nearly uniform distribution on $[0,1]$. It can be done by transforming variables with CDF (cumulative probability distribution) of approximated distribution of this variable, for which we will use Laplace distributions as it agrees well with empirical CDF (Fig. \ref{norm})

Taking $d$ such normalized variables, e.g. for different parameters in given or neighboring times, if uncorrelated they would come from nearly uniform $\rho\approx 1$ distribution on $[0,1]^d$. We will model perturbation from this uniform density as linear combination of orthonormal polynomials $\rho(\textbf{x})=\sum_{\textbf{j}} a_{\textbf{j}} f_{\textbf{j}} (\textbf{x})$. It makes MSE optimal estimation very inexpensive~\cite{me1}: $a_{\textbf{j}}=\frac{1}{|X|} \sum_{\textbf{x}\in X} f_\textbf{j}(\textbf{x})$ is just average over sample $X$. Coefficients for different $\textbf{j}$ are independent and have multivariate cumulant-like specific interpretation, can be used for describing statistical dependencies between tested variables.

 This article extends methodology from \cite{me3} for 1D variable (on example of Dow Jones Industrial Averages time series) into the case of multidimensional random variables.

\section{Normalization to nearly uniform density}
We will discuss on example of time series of 6470 (from 1993 to 2018) daily Yield Curve $\beta_1, \beta_2, \beta_3$ parameters $\{\beta_1,\beta_2,\beta_3\}_{t=1..n_0}$ for $n_0=6470$.

Time series are usually normalized for example to allow assumption of stationary process: such that joint probability distribution does not change while shifting position. The standard approach, especially for Gaussian distribution, is to subtract mean value, then divide by the standard deviation. However, such normalization does not exploit local dependencies between values, what we are interested in.

Hence we will work on sequence of differences (errors, residues) from current value to its prediction based on previous values, which can be taken for example from ARIMA-like models. For simplicity we will use here the previous value as predictor: operate on $\beta_i(t+1)-\beta_i(t)$ sequence for $t=1\ldots n_1$ where $n_1=n_0-1$. In practical applications $\beta_i(t)$ can be replaced with a more sophisticated predictior, for example exploiting long-range dependencies.

As shown in Fig. \ref{norm}, such sequences of differences from predictor turns out to have nearly Laplace distribution:
\be g(y)=\frac{1}{2b}\exp\left(-\frac{|y-\mu|}{b}\right) \label{trans}\ee
where maximum likelihood estimation of parameters is just: $\mu=$ median of $y$, $b=$ mean of $|y-\mu|$.

For simplicity we use Laplace distributions here to normalize variables to nearly uniform in $[0,1]$, with separate parameters for different variables:
\be x_i(t):=G_i(\beta_i(t+1)-\beta_i(t)) \ee
where  $ G(y)=\int_{-\infty}^y g(y')\, dy'$ is CDF of used distribution (Laplace here).

We will search for $\rho_X (x)$ density. To remove transformation (\ref{trans}) to retrieve the final density of $(\beta_1,\beta_2,\beta_3)$, observe that $P(y'=G^{-1}(x) \leq y)=P(x \leq G(y))$. Differentiating over $y$, we get $ \rho_Y (y) = \rho_X(G(y))\cdot g(y)$.

\section{Hierarchical correlation reconstruction}
After normalization we have $\{x_1(t), x_2(t), x_3(t)\}$ time series with nearly uniform density of separate variables. Taking its $d$ values: as different coordinates or neighboring in time, if uncorrelated they would come from nearly uniform distribution in $[0,1]^d$ - difference from uniform distribution describes statistical dependencies in our time series. We will use polynomial to describe this difference: estimate joint density for $d$ neighboring values of $x$. 

Assuming we have $\{\textbf{x}^t\}_{t=1,\ldots,n}\subset[0,1]^d$ vector sequence of neighboring values (we will discuss various possibilities later), we would like to model density of such vectors as polynomial. It turns out~\cite{me1} that using orthonormal basis, which for multidimensional case can be products of 1D orthonormal polynomials, mean square (MSE, $L^2$) optimization leads to extremely simple formula for estimated coefficients:
$$ \rho(\textbf{x})=\sum_{\textbf{j}\in\{0,m\}^d}a_\textbf{j} f_\textbf{j}(\textbf{x})=\sum_{j_1\ldots j_d=0}^m a_\textbf{j}\,f_{j_1}(x_1)\cdot \ldots\cdot f_{j_d}(x_d) $$
\be \textrm{with estimated coefficients:}\quad a_\textbf{j} = \frac{1}{n} \sum_{t=1}^n f_\textbf{j}(\textbf{x}^t) \ee

The basis used this way has $|B|=(m+1)^d$ functions. Beside inexpensive calculation, this simple approach has also very convenient property of coefficients being independent, giving each $\textbf{j}$ unique value and interpretation. Independence also allows for flexibility of considered basis - instead of considering all $\textbf{j}$, we can focus on more promising ones: for example with larger absolute value of coefficient, replacing negligible $a_\textbf{j}$. Instead of MSE optimization, we can use often preferred: likelihood maximization~\cite{me2}, but it requires additional iterative optimization and introduces dependencies between coefficients.

\begin{figure}[t!]
    \centering
        \includegraphics{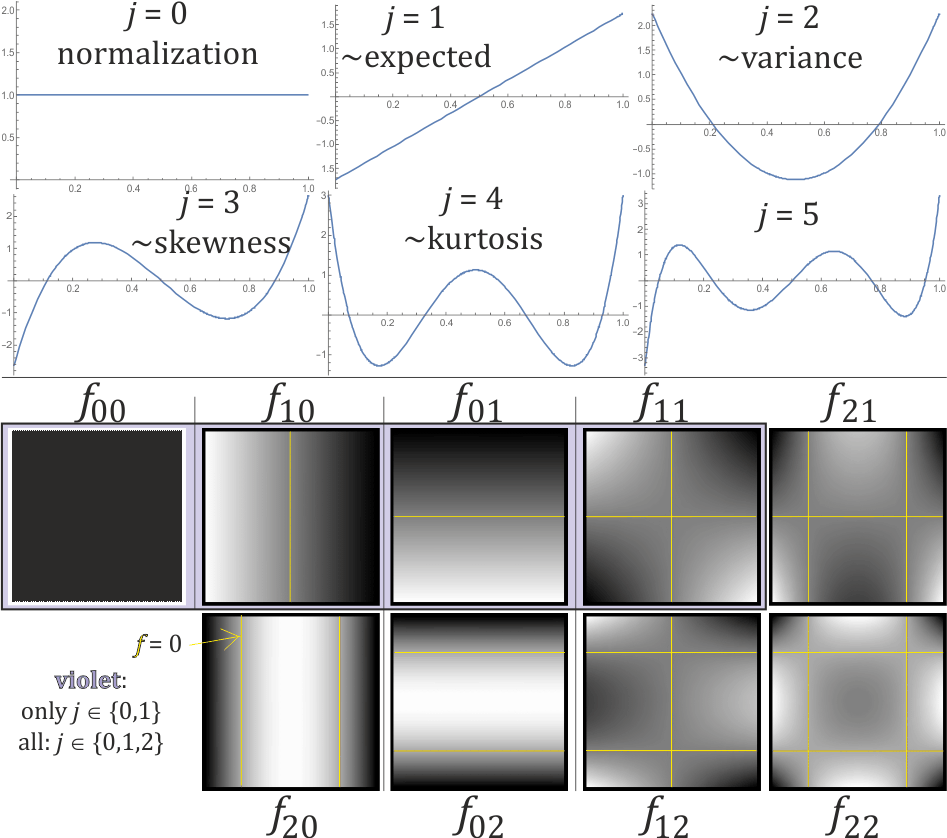}
        \caption{Top: the first 6 of used 1D orthonormal basis of polynomials $(\langle f,g\rangle=\int_0^1 f g\, dx)$: $j=0$ coefficient guards normalization, the remaining functions integrate to 0, and their coefficients describe perturbation from uniform distribution. These coefficients have similar interpretation as cumulants, but are much more convenient for reconstruction of density. Bottom: 2D product basis $f_{\textbf{j}}(\textbf{x})=f_{j_1}(x_1)f_{j_2}(x_2)$ for $m=2$: $j\in \{0,1,2\}$. The $j=0$ coordinates do not modify corresponding variable - generally given coefficient describes statistical dependencies between coordinates having nonzero index.}
       \label{coef}
\end{figure}

Above $f_j$ 1D polynomials are orthonormal in $[0,1]$: $\int_0^1 f_j(x) f_{k}(x) dx = \delta_{jk}$, getting (rescaled Legendre): $f_0=1$ and for $j=1,2,3,4,5$ correspondingly:
$$\sqrt{3}(2x-1), \sqrt{5}(6x^2-6x+1), \sqrt{7}(20x^3-30x^2+12x-1),$$
$$3(70x^4-140x^3+90x^2-20x+1), $$
$$\sqrt{11}(252x^5-630x^4+560x^3-210x^2+30x-1).$$

They are plotted in the top of Fig. \ref{coef}. $f_0$ corresponds to normalization. The $j=1$ coefficient decides about reducing or increasing the mean - have similar interpretation as expected value. Analogously $j=2$ coefficient decides about focusing or spreading given variable, similarly as variance. And so on: further $f_j$ have similar interpretation as cumulants, however, while reconstructing density from moments is a difficult moment problem, presented description is directly coefficients of polynomial estimating the density.

For multiple variables, $a_\textbf{j}$ describes only correlations between $C=\{i:j_i>0\}$ coordinates, does not affect $j_i=0$ coordinates, as we can see in the bottom of Fig. \ref{coef}. Each coefficient has also a specific interpretations here, for example $a_{11}$ decides between increase and decrease of second variable with increase of the first, $a_{12}$ analogously decides focus or spread of the second variable.

Errors of such estimated coefficients come from approximately Gaussian distribution:
\be \tilde{a}_\textbf{j}-a_\textbf{j} \sim \mathcal{N}\left(0,\frac{1}{\sqrt{n}}\sqrt{\int (f_\textbf{j}-a_\textbf{j})^2 \rho\, d\textbf{x}}\right) \label{error}\ee
For $\rho=1$ the integral has value 1, getting $\sigma=1/\sqrt{n}\approx 0.013$ in our case. As we can see in Fig. \ref{contextfree}, many coefficients are more that tenfold larger here: can be considered as essential, not a result of a noise.

\section{Context-free modeling}
\begin{figure}[t!]
    \centering
        \includegraphics{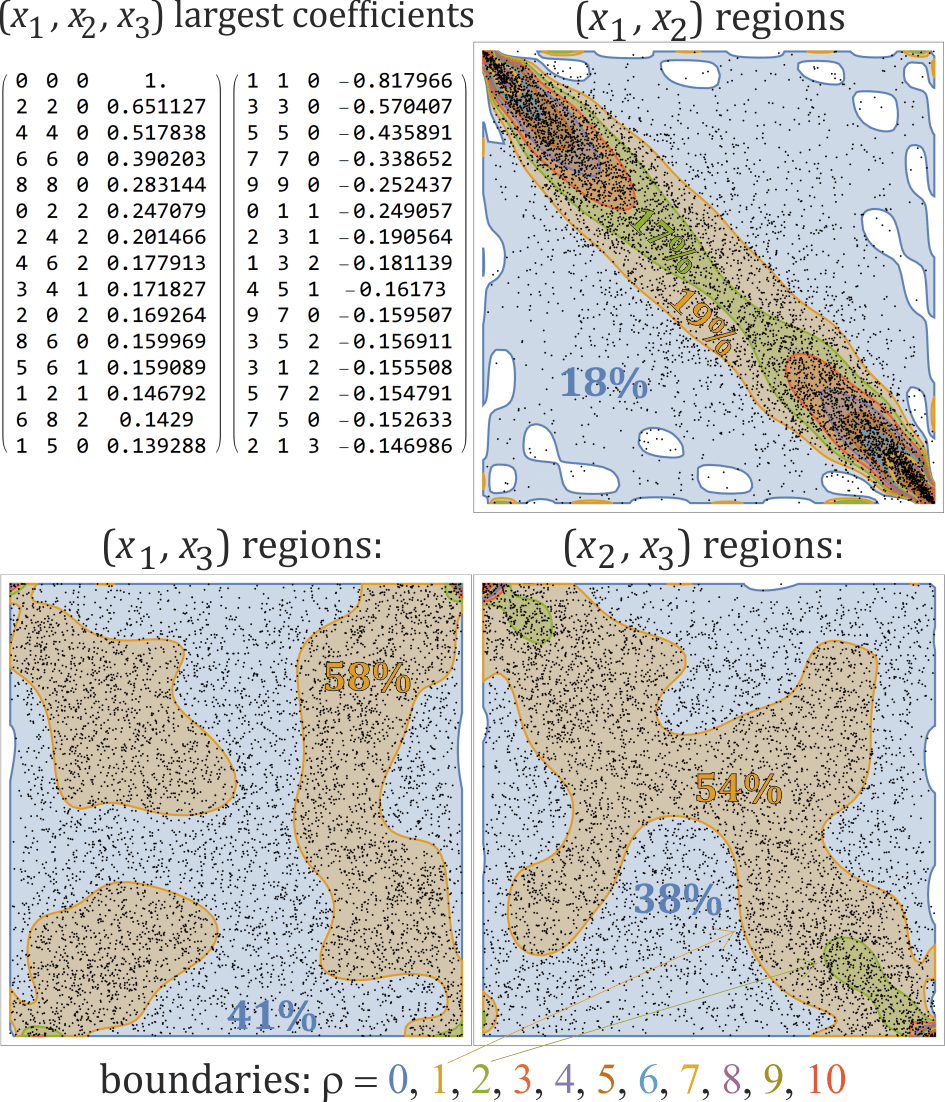}
        \caption{Modelling joint probability distribution of $(x_1, x_2, x_3)$ variables, each normalized to nearly uniform distribution on $[0,1]$. Top left: largest positive and negative obtained coefficients $a_\textbf{j}$ of polynomial used as density estimation: corresponding to $a_\textbf{j}f_{j_1}(x_1)f_{j_2}(x_2)f_{j_3}(x_3)$ correction from uniform distribution on $[0,1]^3$. The remaining 3 region plots show the actual values (6469 tiny black points) and region plots of obtained density as degree $m=9$ polynomial for all 3 pairs of variables, presenting non-uniformity of their joint distribution, especially for the $(x_1,x_2)$ pair (top right). If these variables would be uncorrelated ($\rho\approx 1$), probability of a region would be proportional to its area. In contrast, the blue region here corresponding to estimated density $1\leq \rho\leq 2$ has more than 1/2 of area, but only $\approx 18\%$ of probability, which is mostly concentrated in the diagonal, near its edges. It can be seen in the largest coefficients: negative 110 gives anti-correlation, positive 220 increases probability of extreme values. The third variable appears much further in coefficient list, what means weaker statistical dependency.}
       \label{contextfree}
\end{figure}

We will work analogously to Markov modelling: model probability distribution of a new value basing on one or a few previous values - referred as context for the prediction. In contrast to standard Markov situation, values we are using are continuous: from $[0,1]$ or $[0,1]^2$ or $[0,1]^3$ here. The number of previous values considered for prediction is referred as the order of model. Order 0 or context-free models values as independent random variables: all from the same probability density. Order 1 is standard Markov process: uses context as one previous value to model probability distribution of the current value. Analogously for higher order models: using a few previous values as context for prediction.

Let us start with basic context-free approach: just model joint distribution of observed values, not looking for statistical dependencies with neighboring values. In discussed example $d=3$ dimension and $\textbf{x}^t=(x_1(t), x_2(t),x_3(t))$ for $t=1\ldots n=n_1$.

We could also use only a subset of such variables, e.g. for $d=2$ we have three possible pairs here: $(x_1,x_2)$, $(x_1,x_3)$ and $(x_2,x_3)$. Reducing to $d=1$ should lead to nearly uniform density due to the used normalization. Imperfection of e.g. assumed Laplace distribution used for this purpose will be corrected while fitting polynomial - in multidimensional case by $\textbf{j}$ coefficients with $j_i=0$ for all but a given coordinate.

Top of Fig. \ref{summ} contains evaluation for fitting $m=9$ degree polynomial (for each variable) and 4 cases: $d=3$ and all 3 pairs $d=2$. It shows sorted $\rho$ for predictions of actually observed values - the higher it is, the better prediction. Beside the used model, efficiency of such prediction strongly relies on objective statistical dependencies between these variables - for example will fail if they are uncorrelated. Surprisingly, we can see from this plot that using just first two variables gives better prediction than for all three here - the third variable is weakly correlated, more strongly with $x_2$ (red plot) than with $x_1$ (green).

Such plots evaluating prediction also allow to calibrate density plots, including interpretation for negative densities being artifact of the presented method: we can see that $\approx 3\%$ of cases here got negative density, hence $\rho>0$ region is expected to indicate the proper value in $\approx 97\%$ of cases.

Analogously we can interpret further lines of these graphs, for example we can see that the blue and orange lines in Fig. \ref{summ} have $\rho>2$ in $\approx 62\%$ of cases. Lower 3D density plot shows this $\rho>2$ region in $[0,1]^3$: while it contains only $\approx 14\%$ of the volume, what would be its probability for uncorrelated variables, it contains here much more: $\approx 62\%$ of points from the sample.

2D regions for multiple isolines of constant $\rho$ for all 3 pairs are presented in Fig. \ref{contextfree}, which alternatively can be obtained as marginalization of 3D density. It gives better visualization of strong statistical dependance between $x_1$ and $x_2$, and much weaker with $x_3$. Percentages indicate probability of cases observed in a given region, for example for $0<\rho<1$ for light blue regions. The missing probability is localized in further regions. Tiny black points are the actual 6469 data points - presented density region plots MSE fit degree $m=9$ polynomial to sum of Dirac deltas in all points of the sample. 

Figure \ref{contextfree} also contains the largest positive coefficients (left, always starts with 1 for normalization), and negative (right). They provide unique independent cumulant-like description of statistical dependencies in modelled sample. For example largest positive is $a_{220}\approx 0.65$, what corresponds to parabola in first and second variable: statistical avoidance of being both near the center. Largest negative is $a_{110}\approx -0.82$, saying that with growth of the first variable, there comes reduction of the second.

\section{Context-dependent modelling}
The next step is trying to exploit statistical dependencies between values neighboring in time: based on context representing the history, for example a few previous values, or extracted crucial information about the past for example in some dimensionality reduction method like PCA: corresponding to the largest eigenvalues of covariance matrix.

For simplicity and reducing dimension we will work on $(x_1, x_2)$ pairs as $x_3$ has much weaker correlation. We have considered one previous pair as the context $(d=4)$: $\textbf{x}^t=(x_1(t),x_2(t),x_1(t-1),x_2(t-1))$, or two previous pairs $(d=6)$: $\textbf{x}^t=(x_1(t),x_2(t),x_1(t-1),x_2(t-1),x_1(t-2),x_2(t-2))$ for $t=1\ldots n$ which is $n_1-1$ or $n_1-2$ correspondingly.

\begin{figure}[t!]
    \centering
        \includegraphics{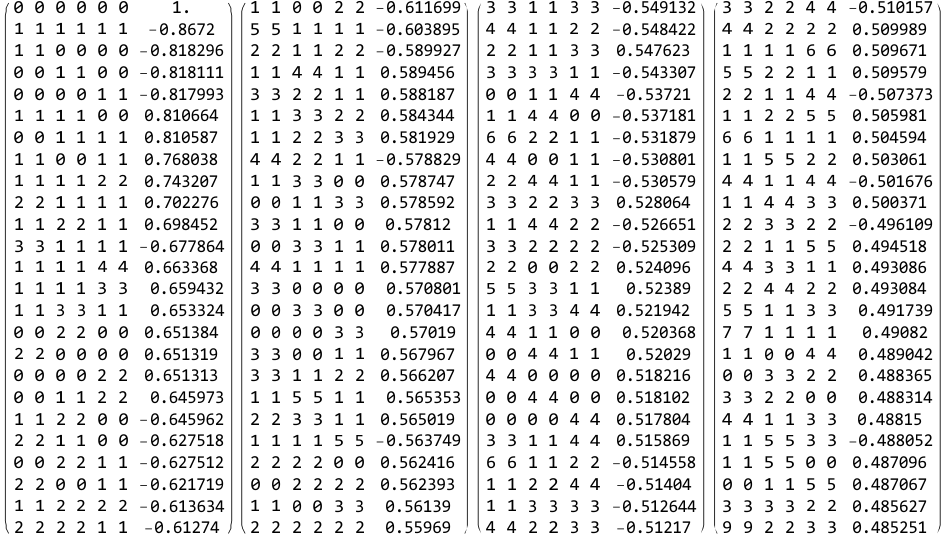}
        \caption{The most significant statistical dependencies: 100 largest absolute value coefficients for the million coefficient model: $m=9$, $d=6$ modeling three neighboring pairs. The corresponding 6 coordinates are: $(x_1(t),x_2(t),x_1(t-1),x_2(t-1),x_1(t-2),x_2(t-2))$.      
        The list obviously starts with $a_{000000}=1$ corresponding to normalization (the remaining functions integrate to 0). Then we have "11" pairs as already seen in Fig. \ref{contextfree}, this time in all 3 positions with nearly identical coefficient (tiny differences come from occurrences at the beginning and the end). Then we see large $a_{111100}\approx a_{001111}\approx 0.81$ positive coefficient describing dependency between neighboring pairs: saying e.g. that with growth of the first 3 variables, the fourth is also likely to grow. While using up to $m=9$ order, we see that the "9" index appears only in the last: 100th position here - dominant statistical dependencies are described by relatively low order polynomials here. Assuming uniform density on $[0,1]^6$, these coefficients should come for Gaussian distribution centered in zero with $\sigma=1/\sqrt{n}\approx 0.012$, hence above coefficients $>40\sigma$ can be seen as statistical significant: should not be interpreted as a result of noise.}
       \label{d6m9coef}
\end{figure}

The most significant 100 coefficients for the largest considered model ($d=6$, $m=9$, $10^6$ coefficients) are presented in Fig. \ref{d6m9coef}. Each is independent and has a specific meaning: correction $a_\textbf{j} \prod_{i=1}^d f_{j_i}(x_i)$ to initially uniform density on $[0,1]^6$ - providing unique description of statistical dependencies in the observed data sample. For certainty that they are not just a result of random noise, $\sigma \approx 0.012$ here for $\rho=1$ (uniform density on $[0,1]^d$), which is exceeded a few dozens of times in this sample.

\begin{figure}[t!]
    \centering
        \includegraphics{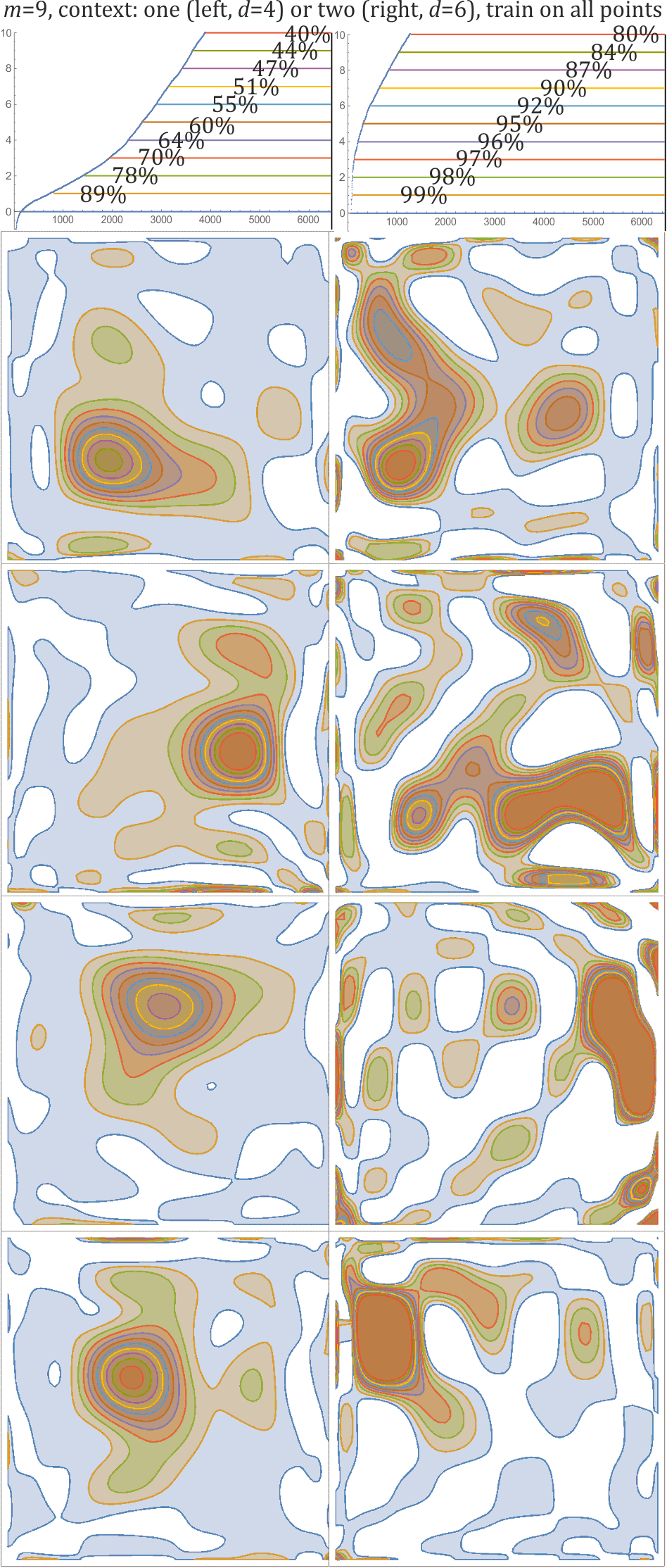}
        \caption{Top: sorted predicted densities for the actually observed values for two degree $m=9$ models: using one (left, $10^4$ coefficients) or two (right, $10^6$ coefficients) previous $(x_1,x_2)$ pairs as the context. It contains percentages of cases when density was above  $\rho=0,1,\ldots,10$ thresholds, drawn below in region plot. Bottom: region plots for predicted densities in some four random points in time, the same for both models. We can see overfitting especially in the right column, with large white regions denoting predicted $\rho<0$. This model fits million coefficients to size 6467 sample - approaching density as polynomial with spikes in the used points. The proper model evaluation should test its generalization capabilities instead: estimate coefficients on a subset of sample, and test on the remaining points - its results are presented in Fig. \ref{contest}. }
       \label{highord}
\end{figure}

The results of this $m=9$ order 2 (right, $10^6$ coefficients) and analogously order 1 (left, $10^4$ coefficients) are presented in Fig. \ref{highord}. Especially the order 2 model gives nearly perfect agreement: in $\approx 80\%$ of cases the actually observed value is in the smallest predicted region (red boundary for $\rho=10$). However, this is fitting million coefficient model to just 6467 data points - polynomial approaching spikes in data points. 

\begin{figure}[t!]
    \centering
        \includegraphics{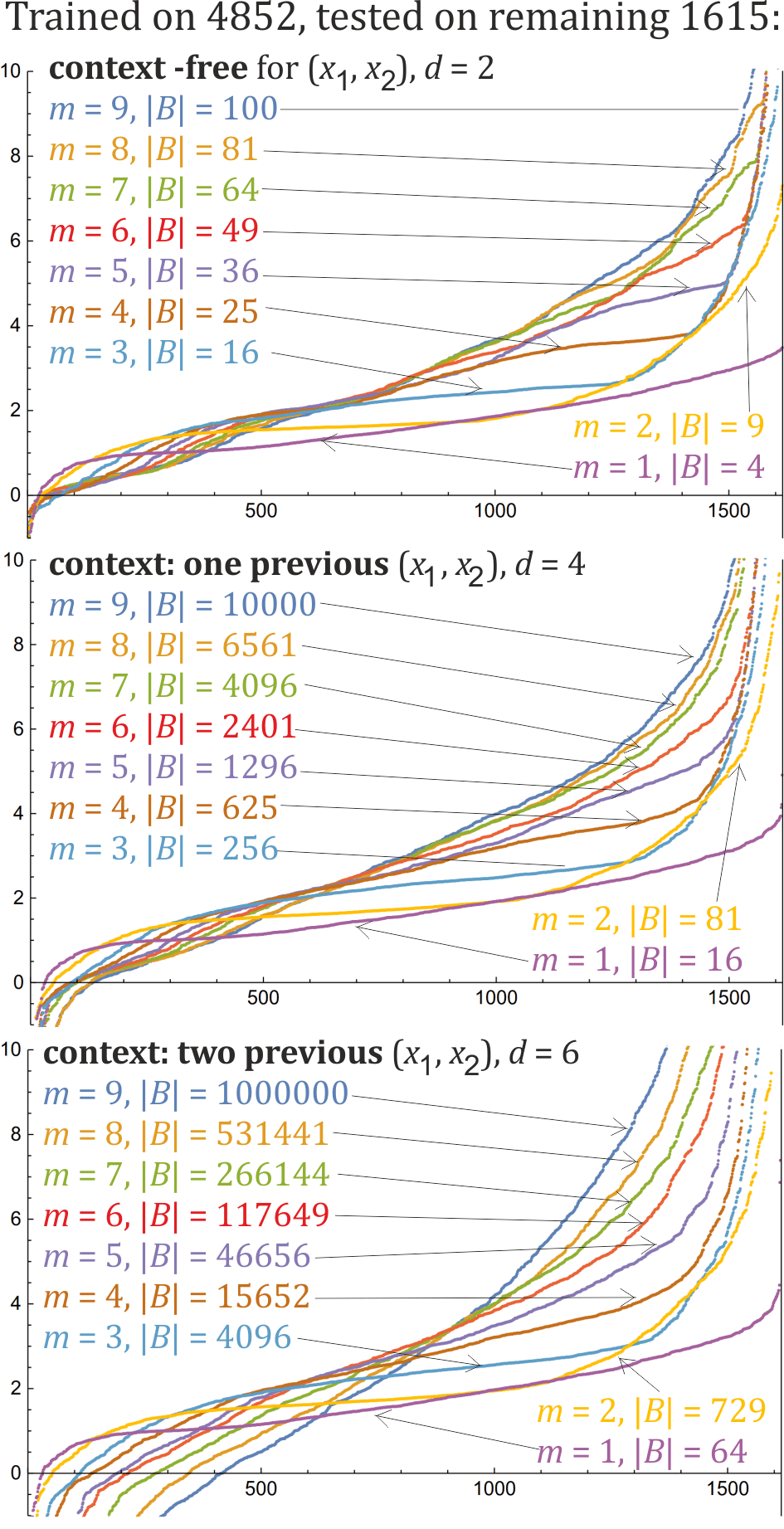}
        \caption{The proper evaluation of 27 models: sorted predicted densities for the actually observed values (the higher the better prediction) in randomly chosen $25\%$ of data points, using the remaining $75\%$ of points to train the model (estimate coefficients). There were used context-free ($d=2$), order 1 ($d=4$) and order 2 ($d=6$) models for $(x_1,x_2)$ and all degrees $m=1,\ldots,9$. The highest (blue) plots are analogous as in Fig. \ref{highord}, but this time with disjoint training and test sets to prevent overfitting.       
        }
       \label{contest}
\end{figure}

The proper prediction evaluation should test generalization capabilities, what is presented in Fig. \ref{contest}. These tests of 27 models first randomly split data sample into two disjoint subsets, use the first one to calculate coefficients, and test on the second subset. We see that the million coefficient model ($d=6, m=9$) in $\approx 25\%$ of cases gives negative density - has strong overfitting. However, focusing on predicted high density regions, it most frequently gives the proper prediction. 

Finally, we see that the choice of the most appropriate model is a difficult question, it might be worth to consider a few models and somehow mix their predictions.

\section{Conclusion and further perspectives}
While there is usually assumed Gaussian distribution for financial data, in reality it is often much more complicated, including multimodal distributions. There was presented basics of systematic approach for modelling such joint distribution with a polynomial - what allows to effectively find and work with parametrisation using thousands of unique and independent cumulant-like coefficients, each one having a specific interpretation, and being inexpensive to calculate.

The used example applied basic methodology for educative reasons, we plan to investigate its extensions in the future, for example:
\begin{itemize}
  \item Selective choice of basis: we have used complete basis of polynomials, what makes its $(m+1)^d$ size impractically large especially for high dimensions. However, usually only a small percentage of coefficients is above noise - we can selectively choose and use a sparse basis of significant values instead - describing real statistical dependencies. Alternatively, we can selectively reduce polynomial degree for some of variables.
  \item Adaptive choice of coefficients: we have assumed that coefficients are constant in time, what corresponds to stationarity of time series. However, in practice it is often non-stationary, what can be modelled using coefficients being not average of all values of a given function like here, but some local averages instead, for example with exponentially decaying weight~\cite{me2}.
  \item Long-range value prediction: combination with state-of-art prediction models exploiting long-range dependencies, for example using a more sophisticated (than just the previous value) predictor of the current value.
  \item Improving information content of context used for prediction: instead of using a few previous values as the context, we can use some features e.g. describing long-range behavior like average over a time window, or for example obtained from dimensionality reduction methods like PCA (principal component analysis).
\end{itemize}
While approach used here was analogous to Markov modelling, alternative approach to consider in the future is using time as one of coordinates, e.g. fit polynomial to $(x_1(t),x_2(t),t)$ triples in moving time window. It would require much lower dimension, allowing to directly model longer correlations. It also allow to work with continuous time.

\bibliographystyle{IEEEtran}
\bibliography{cites}
\end{document}